# ARCHITECTURE AND DESIGN OF MEDICAL PROCESSOR UNITS FOR MEDICAL NETWORKS

Syed V. Ahamed and Syed (Shawon) M. Rahman

Department of Computer Science and Engineering, University of Hawaii-Hilo
Hilo, HI 96720, USA
Email: profahamed@gmail.com and SRahman@hawaii.edu

## ABSTRACT

*This paper[1] introduces analogical and deductive methodologies for the design medical processor units (MPUs). From the study of evolution of numerous earlier processors, we derive the basis for the architecture of MPUs. These specialized processors perform unique medical functions encoded as medical operational codes (mopcs). From a pragmatic perspective, MPUs function very close to CPUs. Both processors have unique operation codes that command the hardware to perform a distinct chain of sub-processes upon operands and generate a specific result unique to the opcode and the operand(s). In medical environments, MPU decodes the mopcs and executes a series of medical sub-processes and sends out secondary commands to the medical machine. Whereas operands in a typical computer system are numerical and logical entities, the operands in medical machine are objects such as such as patients, blood samples, tissues, operating rooms, medical staff, medical bills, patient payments, etc. We follow the functional overlap between the two processes and evolve the design of medical computer systems and networks.*

## 1. INTRODUCTION

The infusion of scientific methodology in the practice of medicine and art of surgery has been a slow but steadily gaining momentum. Tools and technology are much more prevalent now in the diagnosis of ailments and curing of chronic conditions now than ever before. The scientific discoveries and innovation have given the modern doctors a firm rationale over the mere blending of skill and experience of the less scientific approaches of medical practitioners of the 1980s and 1990s. Science is as much a part of modern medicine as aerospace technology was a part of the aircraft industry during the 1980s. Yet the infusion of science and technology from the National Aeronautics and Space Agency (NASA) has substantially transformed the design and manufacture of aircraft and the future of the spacecraft industry.

In the medical field, the first wave of infusion of science and technology has resulted in superb tools. Streamlined procedures and scientific methodologies in the diagnosis of disease and disorder and the subsequent cure of patients is the norm of medical activity. At the current snapshot of time, the infusion has dissipated and the process is now waning into diffusion. In perspective, the second wave of infusion that is still to follow is from the major contributions in processing of information and knowledge. It is slowly and surely replacing processing of numbers

---

[1] This work is partially supported by EPSCoR award EPS-0903833 from the National Science Foundation (NSF) to the University of Hawaii, USA

and data structures in the computer field. Machines are taking over human functions at an unprecedented rate. Learning from human reactions and byelaws of rational behavior, humanistic machines can outperform any unskilled labor force. In these new machines, local reasoning is infused with global wisdom and the immediate outcome is truly beautiful as much as it is socially valuable.

In this paper, we take the bold step of blending the steps in processing of knowledge and information with to the effectiveness and dispersion of global medicinal/surgical practices. The outcome should be a blend of efficiency and accuracy of digital techniques in computers and networks with the strides of scientific discoveries and innovations in the field of modern medicine. Further, in this paper, the basis of modern medicine is assumed to be in some documented form of computer and network accessible (textual, graphical, accounting, observational, statistical, etc.) files. Opinions and experience of the professionals are summarized as derived expert system knowledge bases and/or computer driven AI generated files that can be used as reference nodes in the deduction and derivation of new hypothesis, conclusion or inference. Information and knowledge processing techniques merge and process the current local conditions and constraints with those observed in the past and documented in the medical knowledge bases around the globe. Two major effects become evident.

*First*, the medical machines provide sound and documented sets of recommended procedural steps to solve the local problem dealing with diagnosis and/or cure. The confidence in each of the derived inference is numerically displayed for the local team of specialists to verify and modify (if necessary) thus forcing the practitioners to keep abreast of global knowledge. The machine process approaches the methodology now in vague for forensics studies and crackdown for the breach of computer security.

*Second*, the medical machine tracks the difference between its prediction based on prior historical facts in the knowledge bases and the observed (or preferred) outcome or treatment. The AI based learning algorithms validate the cause-effect relationship in the current/local problem and logs the findings for more human (humanistic) consideration.

Both actions of the medical machine make its functions complimentary and subservient to the surveillance of human judgment. Major modification to the medical and procedural knowledge bases is completed only under the consensus and validation of global medical experts. Such practices are frequently undertaken in most national and international knowledge bases in any given discipline (such as economics, finance, banking, economics, or any expertise) only if there is agreement and consensus among the experts in that particular discipline.

## 1.1 MEDICAL KNOWLEDGE BASES

The contributions to systematic accumulation of knowledge pertaining to medical sciences have been scattered and diversified. In a majority of cases the activity has been commercial rather than academic and scholarly. The computational and network tools and techniques have been obscured by the clout of profiteering from the sale of data to specialized medical teams and desperate patients. As late as 2010, only a few nonprofit medical knowledge bases have emerged. The authenticity and accuracy of the contents is by entry rather than by reason and scientific accountability. Three key components to deploy knowledge processing techniques in global medical networks are *(a)* the Internet accessible knowledge bases that hold, share and validate the current medical knowledge and procedures in every well-defined area of expertise and information, *(b)* highly efficient MPUs to analyze and resolve local medical issues and distribute the procedural commands to the medical computers, their peripherals, networks and knowledge bases and *(c)* highly evolved and efficient backbone and global networks. Unified, integrated and streamlined, these three components constitute the next generation medical networks

The NIH funded research at Stanford University[2] has yielded noteworthy contributions such as Internist-I [14] as far back as 1982. The ensuing GUIDON (an intelligent computer aided instruction tool)[8], and MYCIN[16], etc. lack the knowledge about structure and the strategy in the diagnosis and the rationality for the expert system based recommendations for procedures and/or therapy. Initially, it was used for determining the recommendations for bacteremia and meningitis. MYCIN has a non psychological, expert-system, probability driven knowledge base primarily used for referencing rather than for analysis, derivation, and enhancement of medical knowledge. The heuristic approach has serious limitation for the creative deployment of medical knowledge for the particular ailment of a particular patient. Patients are unique and highly individualistic; only medical knowledge is generic and the role of the medical machine is to compliment the generic knowledge with unique circumstance and individuality of the patient. Conversely, the medical machine actively deploys the scientific reasoning and methodology to suit the unique needs of patient. The medical team monitors the reason and logic that the medical machine is currently using to fine tune the outcome to be (pretty much) flawless and meets the desired goal within a present level of confidence. Surely, it may not circumvent the inevitable, but it will leave no stone unturned (in the world wide medical knowledge domain) to delay or even bypass the inevitable for long spans of time.

Initial medical procedural codes, their associated details, ensuing methodology and their databases were initiated at the City University of New York as far back as 1990 by Krol[12]. Integrated medical network (IMN) configurations and their architectures have been studied by numerous other researchers such as Mollah[15] and Kazmi[11]. A novel encoding of the procedural codes in ATM cell was proposed by Lueng [13]. Waraporn introduced graphing of symptoms and deduction ailments and their treatments in the medical field in his 2006 dissertation[21] and the following publications[5][22].

The concept of multidimensional Hamming codes was introduced in 2007[1] [3] to the medical field in order to identify the nearest set of documented symptoms of diagnosed ailments to those of the patient. Being programmable these steps facilitate quick, efficient and quantifiable results in the diagnosis and then the treatment of the patient condition by matching the personal history and data of the patient with those of successfully treated patients. Medical knowledge bases are an integral part of the studies by Mollah and Waraporn.

In order to comprehensively integrate the artificial intelligence techniques in the Web environment, Ahamed has extended this methodology[1] over the last few years. Rahman[19] has also documented the most recent finding through 2010. Numerous configurations that utilize the intelligent Internet architectures and the national medical expert teams are presented in Section 3. These intelligent medical systems environments, based on Internet access constantly scan the activities in hospitals and medical centers in order to filter out the routine activity and select the unusual circumstances to add to the medical knowledge bases around the country or the world.

---

[2] The long list of contributions from NIH funded Stanford University's scientists is remarkable. In particular, Internist, Mycin, Mycin-II and NeoMycin have transformed the methodologies for merging heuristic programming into the practice of medicine.

In a recent IBM patent Basson, et., al.[7], have indicated the crucial use of sensing techniques to develop a vehicle control system when the driver suffers from potentially debilitating conditions. A database including such conditions and symptoms is accessed. The condition is then processed by a traditional CPU with network access to central control and attempts to rescue the situation and also the medical emergencies for the drive. In an earlier paper [18] highlights the concept of an "intelligent" garment worn by human beings that monitors the Georgia Tech Wearable Motherboard (GTWM) to monitor the vital signs in an unobtrusive manner. The sensing of EKGs, temperature, voice recorders, etc., and permits the wearer to transmit data and information to external devices. In 2006, Doukas, et al.,[9] have proposed advanced telemedicine services the context-aware networks.

## 2. PROCESSOR DESIGNS

### 2.1. Platform of Current Processors

Medical processor unit (MPU) designs are firmly entrenched in the design of processors. central processor units[3], typically, (CPUs), numerical processor units (NPUs), digital signal processors (DSPs) and knowledge processor units (KPUs). In the section, we briefly review the architectures of the simplest graphical processor unit (based on the vintage von Nuemann CPU) design to the more elaborate knowledge processor unit (KPU) design.

Central processor units have been the generic hardware for numeric, arithmetic, and logic functions since the von Neumann machine [10] in 1945. The architecture of the IAS (Institute of Advanced Study) machine [17] was the first blue print of many more machines that were built for a decade or so after the basis of the IAS machine was well received by the computer manufacturers. The earlier CPU architectures have described in detail [3].

### 2.2. Display/Graphics Processor Units (DPUs/GPUs)

The earlier display devices (DPUs) for cathode ray tube (CRT) technology are obsolete. Cursor input in the CRTs is very outdated. The software was primitive and the display functions and colors needed considerable enhancements. The flat panel technology and the availability of very fine granularity and colors options have rendered new venues for machines to carry the visual impact deep in the minds and thoughts of scientists and researchers alike. The impact in every scientific and social direction of progress is monumental.

During their introduction, the GPUs served as the intermediate processors between the CPU and the graphic devices. The GPUs were initially built to relieve the CPU from the display functions and continue with the application based processing. Earlier designs (e.g., Intel 82786 graphics coprocessor) adapted a separate configuration but the later CPU architectures absorbed the basic GPU configurations within the larger VLSI chips that were offering larger capacity, dependability, and yield. Typical configuration of a CPU from the 1980s, that could also be deployed as a GPU is shown in Figure 1. The GPUs for movie industry are far more specialized and sophisticated.

---

[3] In the modern days, the number of processors has increased dramatically. It is possible to build a processor for almost any function or task, ranging from arithmetic functions to automatic robotic responses. In this section we present the trail of only a few pertinent processors that form a basis for the evolution of the MPUs.

The internal bus structure within the GPU is based on the word-length and the instruction set of the graphics/display processor unit. Typically, the older (1980s and early 1990s era) graphics processors and accelerators (the Intel 80860, i860, 1989) delivered 13 MFLOPS or approximately half million-vector transformations for graphics applications. The architecture deployed 10 bus structures, integer and floating-point adders and multipliers, 2 cache memories, and numerous switches and multiplexer units. The algorithms for computer graphics functions are highly optimized to deliver the most desirable processor power consistent with desired level of visual impact to the viewers based on the application requirements.

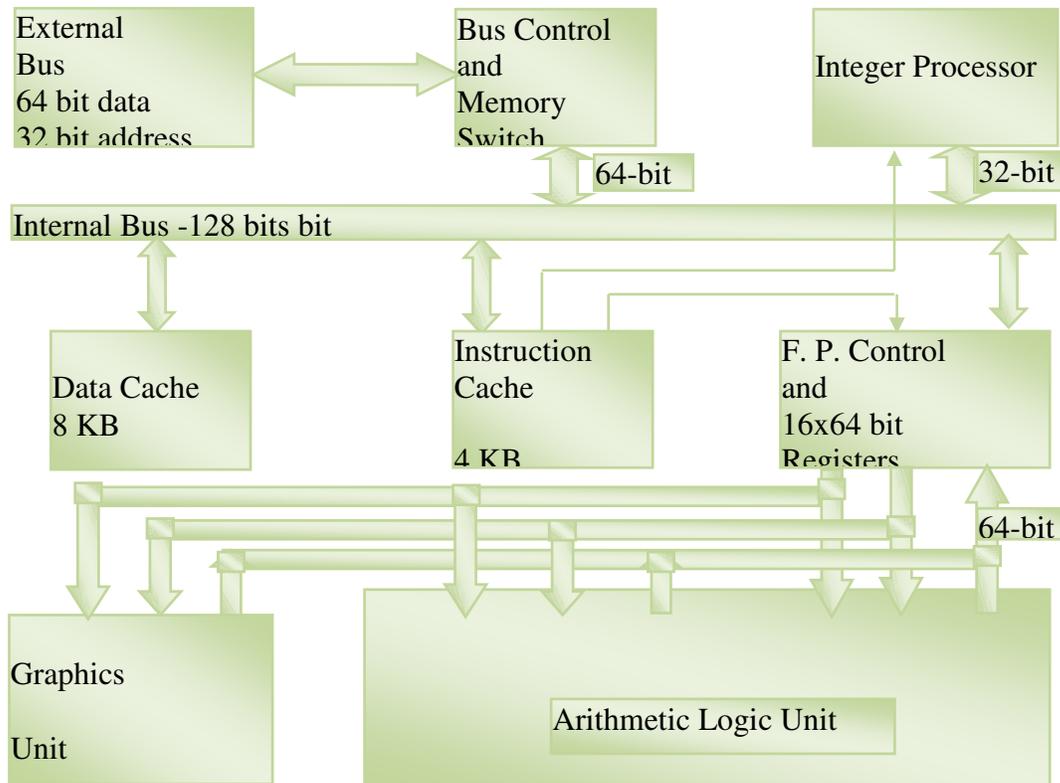

Figure 1  A typical CPU configuration that could be used as a GPU unit form the late 1980 era.

Recent graphics processors are as powerful as the CPUs.  In fact, the VLSI technology is the same and some GPUs outperform traditional CPUs for specific applications. Movie making has initiated a whole new era of graphical computer systems.  Processors for graphical inputs and outputs have specialized requirements.  Lapse-time and real-time processing are both prevalent. In real-time applications the GPUs keep pace with the application-based computations.  When the entire systems are designed for special applications, (such as distance learning, remote surgery, MRI, stock market, games, etc.) the response time of the both the two (application oriented, and graphics oriented) subsystems need consideration. The communication capacity is influenced on channel capacity and the three (computation, graphics, and communication) basic design parameters influence the accuracy, clarity, and quality of the output displays, movies and pictures.

Graphical user interfaces (GUI) provide the user inputs to applications programs that generate the resulting displays. Graphical devices are main output devices for computer to communicate with the users. In certain applications such as computer games and movie making, the perception time of the users and viewer becomes crucial. When the computer systems are dedicated to movie making, the graphics processing systems can become as elaborate and intricate as mainframe computers. The balance between the processing power and graphics display capability becomes an option for most of the designers of standard desktop and laptop computers. In the customized systems, the processing powers are based on the choice of the CPU, GPU processor chips and the *opcodes* that drive them.

For intensive graphics application independent graphics processor are deployed. Most modern computing systems have integrated GPUs built in the overall architecture. GPU computing has become a specialty in its own right. It is seen as a viable alternative to the traditional microprocessor based computing in high-performance computing environments. When used in conjunction with selected CPU's, an order of magnitude gain in performance can be expected in game physics and in computational biophysics applications. The parallel architecture of the GPU is generally used to accelerate the computational performance. These GPUs handle a broad range of process intensive complex problems with a visual insight into the nature of changes that occur as the computation progresses.

## 2.3. Object Processor Units (OPUs)

The architectural framework of typical object processor units (OPUs) is consistent with the typical representation of CPUs. Design of the object operation code (*oopc*) plays an important role in the design of OPU and object oriented machine. In an elementary sense, this role is comparable to role of the 8-bit operation code (*opc*) in the design of IAS machine during the 1944-45 time-frames. For this (IAS) machine the *opc* length was 8 bits in the 20-bit instructions, and the memory of 4096 word, 40-bit memory corresponds to the address space of 12 binary bits. The design experience of the game-processors and the modern graphical processor units will serve as a platform for the design of the OPUs and hardware based object machines.

The intermediate generations of machines, (such as IBM 7094, 360-series) provide a rich array of guidelines to derive the instruction sets for the OPUs. If a set of object-registers, or an object-cache can be envisioned in the OPU, then the instructions corresponding to Register instructions, (R-series), Register-Storage (RS-series), Storage (SS), Immediate operand (I-series), I/O series instructions for OPU can also be designed. The instruction-set will need an expansion to suit the application. It is logical to foresee the need of control-object-memories to replace the control memories of the micro-programmable computers.

The instruction set of the OPU is derived from the most frequent object functions such as (a) single object instructions, (b) multiple object instructions, (c) object to object-memory instructions, (d) internal object-external object instructions, and (e) object-relationship instructions. The separation of logical, numeric, semi-numeric, alphanumeric, and convolutions functions between objects will also be necessary. Hardware, firmware, or brute-force software (compiler power) can accomplish these functions. The need for the next-generation object and knowledge machines (discussed in Section III) should provide an economic incentive to develop these architectural improvements beyond the basic OPU configuration shown in Figure 2. The designs of OPU can be as diversified as the designs of a CPU. The CPU's, I/O device interfaces, different memory units, and direct memory access HW units for high-speed data exchange between main memory units and large secondary memories. Over the decades, numerous CPU architectures (single bus, multi-bus, hardwired, micro- and nano-programmed, multiple control memory based systems) have come and gone. Some of micro-programmable and RISC architecture still exist. Efficient and optimal performance from the CPUs need also combined SISD, SIMD, MISD, MIMD, [20] and/or pipeline architectures. Combined CPU designs can use different clusters of architecture for their sub-functions. Some formats (e.g., array processors, matrix manipulators, etc.) are in active use. Two concepts that have survived many generations of CPUs are *(a)* the algebra of functions (i.e., *opcodes*) that is well delineated, accepted and documented and *(b)* the operands that undergo dynamic changes as the *opcode* is executed in the CPU(s).

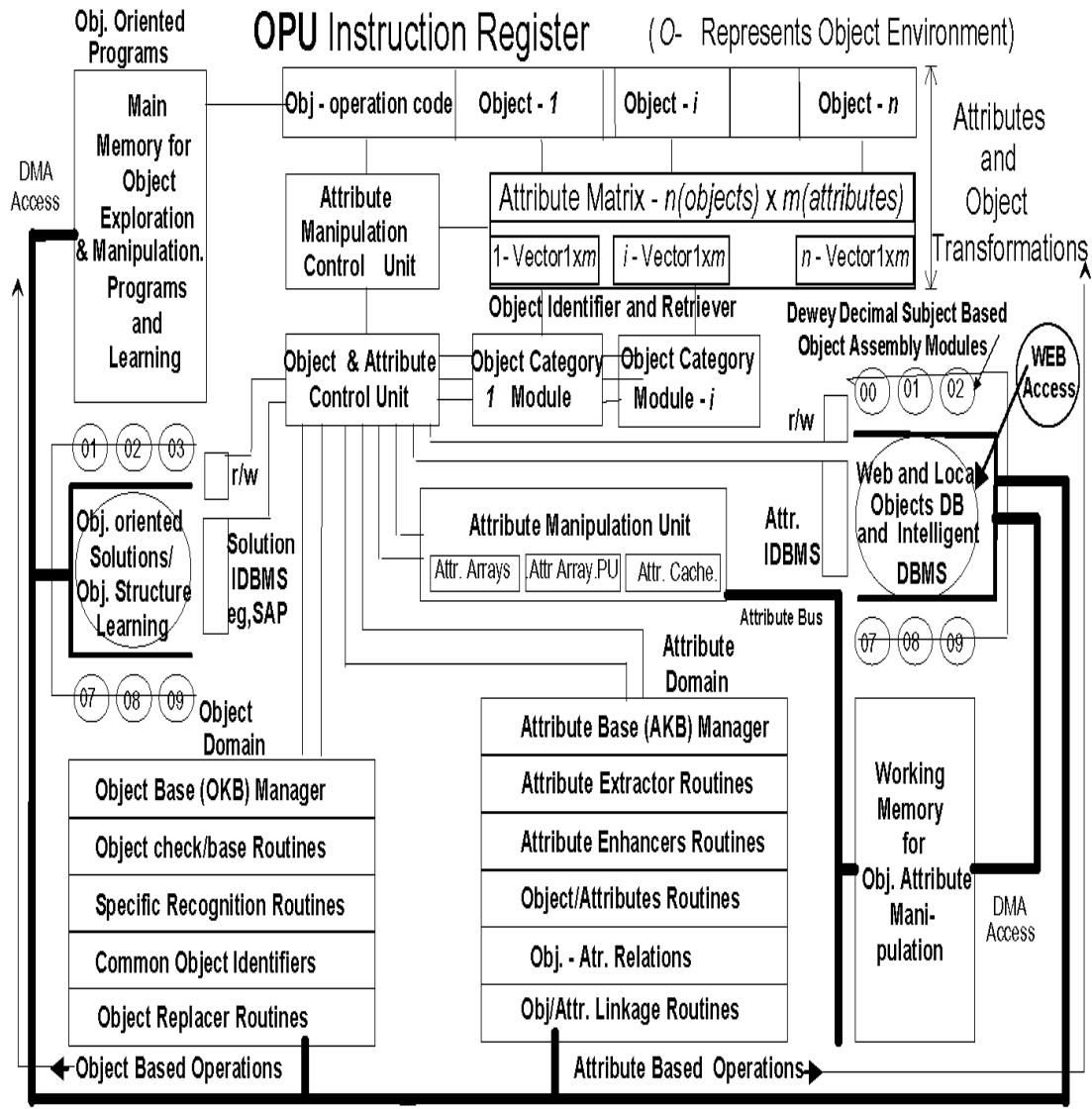

Figure 2. Schematic of a hardwired object processor unit (OPU). Processing n objects with m (maximum) attributes generates a n×m matrix. The common, interactive, and overlapping attributes are thus reconfigured to establish primary and secondary relationships between objects. DMA = direct memory access, IDBMS = Intelligent, data, object and attribute base(s) management system(s), KB = knowledge base(s). Many variations can be derived.

An architectural consonance exists between CPUs and OPUs. In pursuing the similarities, the five variations (SISD, SIMD, MISD, MIMD, and/or pipeline) design established for CPUs can be mapped into five corresponding designs; single process single object (SPSO), single process multiple objects (SPMO), multiple process single object (MPSO), multiple process multiple objects (MPMO), and/or fractional process pipeline respectively [4].

## 2.4. Knowledge Processor Units (KPUs)

Knowledge is derived from objects, their nature, attributes, and their interactions. Thus, the processing capability of knowledge entails processing objects, their attributes and object interactions. Numerous design of KPUs become evident and in fact, they can be derived from the varieties of CPU's initially and then the GPUs that can function as OPUs, and the finally the GPU's that can also serve as CPU's. The creativity of the individual KPU designer lies in matching the HW architecture to the application needs. KPU chips being more expensive and process intensive than CPUs, are unlikely to become as numerous as CPUs that can be personalized to any whim and fancy of the chip manufacturers. The function of the KPUs depends on the capacity of the HW to manipulate or juggle (global and local) objects, based on their own syntax, and environmental constraints in the semantics of the user objective.

The CPU's functionality depends on the capacity to execute stylized operation codes on arithmetic and logical operands (in highly specialized formats and date structures). The configuration of a simple KPU operating of object operands is shown in Figure 3. Other variations based on SIMD, MISD, and MIMD[4] and pipeline architectures of the CPU have been built and such variations can be transfused in the OPU designs. Object processors that lie in between CPUs and KPUs bring in another degree of freedom because KPUs can deploy OPUs, much like CPU's can deploy ALUs and NPUs. Sequential, pipeline, and parallel execution of operations on objects in KPUs gives rise to at least eight possibilities; SKI-SO processors, SKI-MO processors, MKI-SO processors, and MKI-MO processors. Now if SO and MO processors have SOI-SO, SOI-MO and MOI-SO (pipeline structure), and MOI-MO (pipeline and/or multi-processor structure) have variation embedded within themselves, then at least eight design variations become evident. The SKI-SOI-SO is the simplest to build while the MKI-MOI-MO is the most cumbersome to build. From the first estimate, the HW for the simplest KPUs should be an order of magnitude more complex the IBM 360 CPU's (even though these CPUs deployed the microcode technology).

Knowledge processing is based on rudimentary yet pragmatic knowledge theory[1]]. Stated simply, human knowledge is clustered around objects and object groups. Such objects can be represented by data and information structures. Data has numerous representations and information has several forms of graphs and relationships that bring order and coherence to the collection of objects. Such a superstructure of data (at the leaf level), objects (at the twig level), the object clusters (at the branch level) constitute a tree of knowledge. Specific graphs moreover, relationships that bind information into a cogent and coherent body of knowledge bring (precedent, antecedent, and descendant) nodal hierarchy in a visual sense that corresponds to reality.

---

[4] A brief explanation of acronyms for this section is presented. CPU = central processor unit; KPU = knowledge processor unit; ALU = Arithmetic logic unit; NPU = numeric processor unit; SIMD = single instruction, multiple data; MISD = multiple instruction, single data; ,MIMD = multiple instruction, multiple data; OPU = object processor unit; SKI-SO = single knowledge instruction-single object; SKI-MO = single knowledge instruction- multiple object; MKI-SO = multiple knowledge instruction- single object; MKI-MO = multiple knowledge instruction- multiple object; SO processor = single object processor; MO processors = multiple object processors; SOI-SO = single object instruction-single object; SOI-MO = single object instruction-multiple objects and MOI-SO = multiple object instruction-single object; MOI-MO = multiple object instruction-multiple object SKI-SOI-SO = single knowledge instruction-single object instruction-single object; MKI-MOI-MO multiple knowledge instruction-multiple object instruction-multiple objects.

Knowledge processor units should be able to prune, build and shape, reshape and optimally reconfigure knowledge trees, much as CPU's are able to perform the arithmetic (and logic) functions on numbers and symbols and derive new numbers (and logical entities) from old numbers (and logical symbols).

In the design considerations of the CPU, the more elaborate AU functions are known to be decomposable into basic integer and floating-point numeric (add, divide, etc.) operations. Similarly complex logical operations can be reconstituted as modular (AND, OR, EXOR, etc.) functions. In this vein, we propose that all the most frequently used knowledge functions are feasible for the KPU to perform the basic, elementary, and modular functions on objects.

Knowledge bearing objects can be arbitrarily complex. Numerous lower level objects can constitute a more elaborate object entity. Like bacterial colonies, knowledge superstructures have dynamic life cycles. The order and methodology in the construction and destruction of such knowledge superstructures leads to "laws of knowledge-physics" in the knowledge domain under the DDS classification 530 to 539 Traditional laws of Boolean algebra and binary arithmetic do not offer sufficient tools for the calculus of the dynamic bodies of knowledge undergoing social and technological forces in society. A brand new science of knowledge would be appropriate for knowledge (humanistic) machines to perform in the human domain as well as computers perform in the numerical domain. Object and knowledge machines would pave the way. If the new laws for the flow, dynamics, velocity, and acceleration of knowledge and knowledge centric objects (KCOs) can be based on a set of orderly, systematic, and realistic knowledge operation codes (*kopcs*), then these laws can be written as machine executable routines that operate on the KCOs. This approach is a bold digression from the approach in classical sciences where the new concepts enter the sciences as symbolic and mathematical equations. In the modern society, information errupts as multimedia WWW streams, rather than as a gracefully expansion of coherent and cogent concepts embedded in knowledge. Time for extensive human contemplations is a rare luxury to reset the origin of knowledge. Much as we needed digital data scanning systems for DSPs in the past, we need a machine-based common sense, sensing systems to separate junk level information [2] from knowledge bearing information. Knowledge filtering [1] accomplishes this initial, robust, sensible, and necessary task.

The current scenario of science and innovation has given rise to the deployment of technology before it gets obsolete. To accommodate this acceleration of knowledge, we propose that we have a standard set of basic and modular *kopcs* [1]. The complex knowledge operations that encompass the newest concepts are then assembled from the basic set of internationally accepted standard of *kopcs*. The proposed representation for the dynamics of knowledge paves the way between concepts that create new knowledge and the technology that uses such knowledge. The international telecommunication union (ITU) effort has standardized efficient protocols for the seven layer OSI model for communication of data, and then again as the TCP/IP protocol.

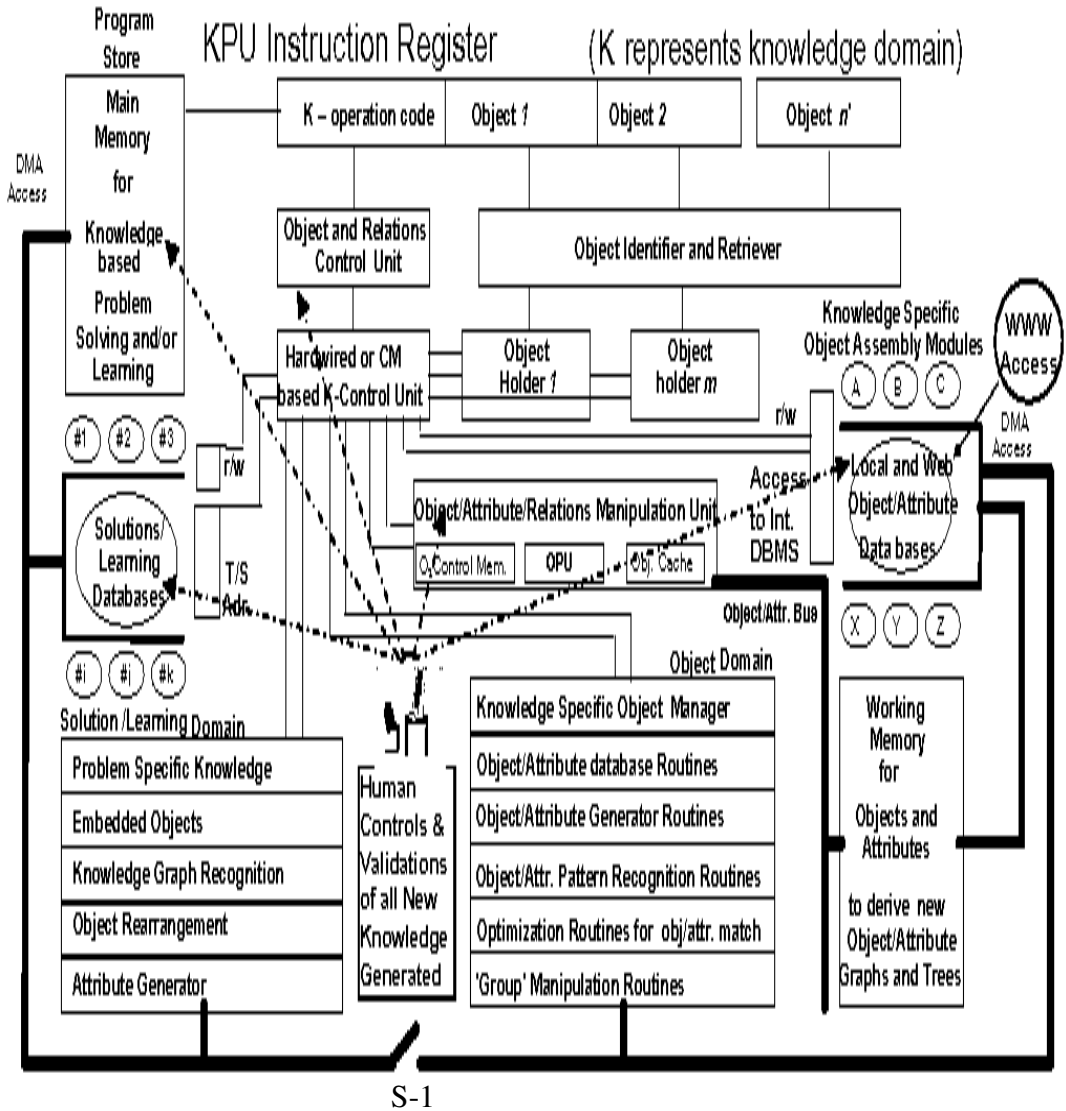

S-1

Figure 3 Switch S-1, Open for Execution Mode for Knowledge Domain Problem Solving; Closed for Learning Mode. The Learning programs 'process' the existing solutions and are able to extract Objects, Groups, Relationships, Opcodes, Group Operators, Modules, Strategies, Optimization Methodologies from existing solutions and store them in Object and corresponding databases. The architecture permits the KPU to catalog a new object in relation to existing objects and generate/modify existing pointers to and from new objects.

## 3. DESIGN OF MEDICAL PROCESSORS

### 3.1. Medical Processor Units (MPUs)

Medical processor units (MPUs) have not yet specifically built for medical operation codes (*mopcs*). It is possible and viable that such processor chips will be tailored and geared to medical functions. A variety of such generic functions can be readily identified. One such medical operation code (*mopc*) is to instruct the MPU to compare bacteria strain (*i*) with all known bacteria (*j, j = 1, n*) from WWW bacteria banks and identify any settle change patterns. Certain amount of object processing and pattern recognition procedures will become essential to complete this task. For computer scientists and VLSI designers the modular tasks involved in executing this overall task have already been accomplished in other disciplines (such as face/fingerprint-recognition in criminology or cell shape recognition in medicine). The ordeal is to assemble an interdisciplinary team of scientists.

Other generic functions are the identification of ailments, patient complaints/cures, most effective treatments, and knowledge processing units to bridge the various knowledge domains. Only a prolonged and consistent effort from the medical staff can lead to generic operation codes for MPUs that can be focal element of medical computer. For an initial step, the knowledge operation codes (*kopcs*) should suffice as medical operation codes (*mopcs*) discussed in this section. Medical processors and medical data banks work in tandem for executing a numerous instructions, retrieving/storing, and processing pertinent medical information. Numerous medical processor hardware units and modules, (including conventional CPUs) coexist in the integrated system to track the confidence levels of the medical functions, individually and collectively. Figure 4 depicts the schematic of a medical processor unit (MPU) derived from the OPU shown in Figure 2. Some of the features for automated learning procedure from the KPU shown in Figure 3 are included for the medical systems to activate AI procedures and suggest improvised procedures to the medical teams. A typical wide area based medical network with advanced learning features and retrieval capability is shown in Figure 5.

Patients and their attributes are treated in unison and as one synergy. Patient records are dynamic and reflect the most recent state of health and the attributes are updated with a history of changes. Procedures are configured from known procedures on similar patient but with slightly different attributes. The procedures are mapped against the attributes to ascertain the optimality and efficacy of treatments and procedures. The basic laws of medical sciences are not violated during administration of treatments and procedures. Administering *n* sub-procedures on a patient with *m* (maximum) attributes generates a $n \times m$ matrix. The common, conflicting and overlapping attributes are thus reconfigured to establish primary and secondary medical safety rules (diabetes and sugar control, alcohol and drugs precaution, etc.) in the realm of medical practice even by a machine. More complex rules are deducible based on medical database (MDB) information for any specific condition/patient.

### 3.2. Medical Computers

Medical computers constitute a special breed of cluster computers. Both have identifiable supersets of traditional computers for the routine functions. While it is viable to force a cluster computer to act as full fledge medical computer, it is desirable to build medical computers on a distinctive track of its own medical operation codes (mopcs). The demand for such medical computers is likely to expand and grow like the specialized air borne computers for the aviation industry.

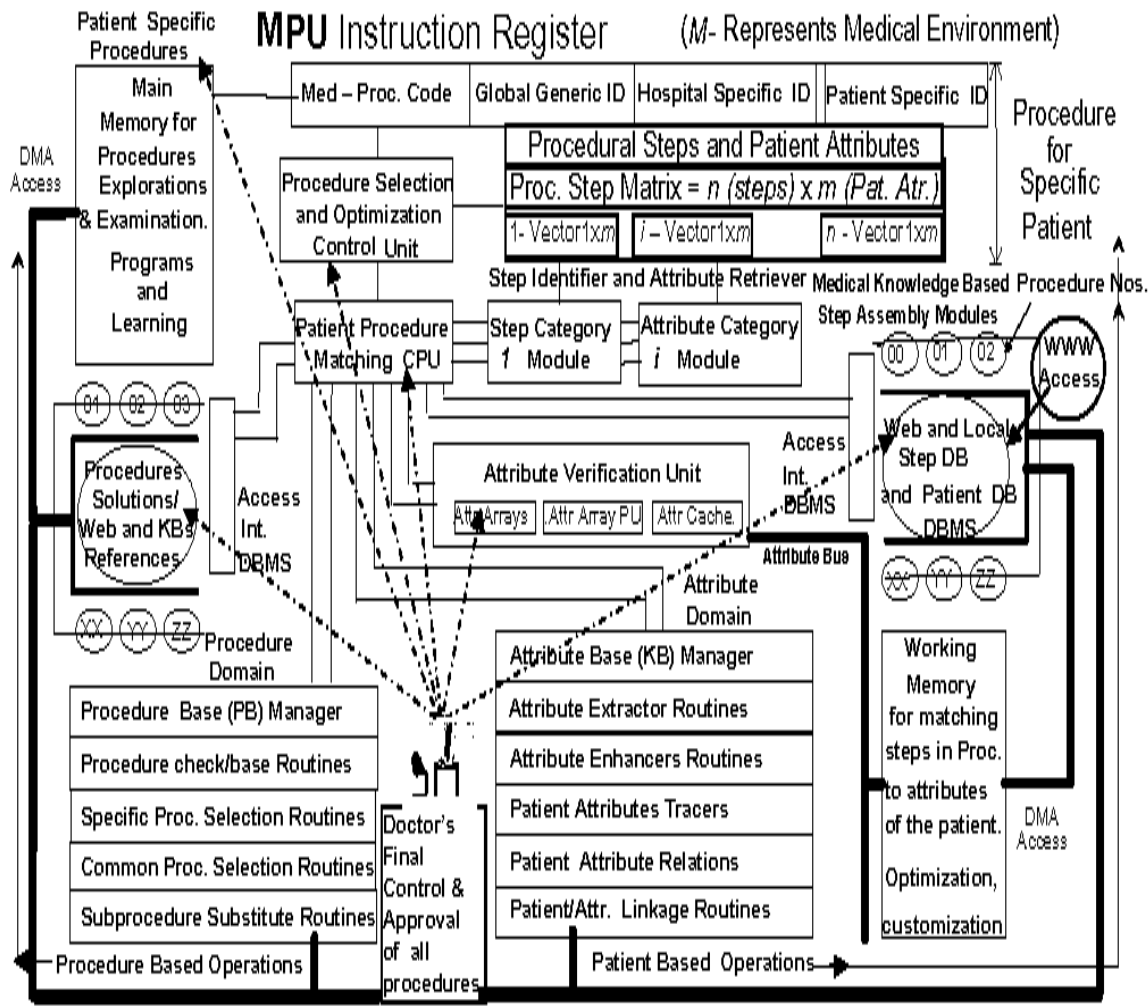

*Figure 4. Architecture of a medical processor unit (MPU). DMA = direct memory access, DBMS = Data, procedure, patient and/or attribute base(s) management system(s), KB = knowledge base(s).*

Patient databases, physician access points, patient access points, and service facilities are connected to the medical data banks and medical processor via several buses. In an alternative integrated medical computer system, numerous processors are included with their own memories and modules and are linked together to establish a processor net unit. Such systems are amenable to hospital-campus settings, where several buildings comprise the hospital, or where several hospitals are interlinked over local area networks.

**3.3 Intelligent Medical Networks**

In the intelligent and integrated medical facilities, numerous MPUs and knowledge banks are linked via a high-speed backbone network as depicted in Figure 5. Isolated packets of information arrive at the knowledge banks from numerous medical centers and hospitals. Optimal protocol design and packet structure for medical functions needs to be evolved for this type of medical environment, even though the existing TCP/IP protocol will suffice. The addressing of the distant knowledge banks is done via a subject matter identifier allocated to the distant knowledge banks. Content and address map of the entire network is housed in specialized medical service "control point" that facilitates[5] the high-speed backbone network to quickly access the information for real-time or emergency conditions.

This identifier of the knowledge bank is consistent with the information stored in that particular bank, thus reducing the switching time to these massive information stores. In such systems, the instruction to the knowledge bank is followed by a burst of input data via the packet switching network. The maintenance of the medical data bases (MDBs in Figure 5) is provided by expert teams who update and track any innovations/changes in the profession. These specialized services are provided by the team of knowledge maintenance systems specialists (see KMS-1 and KMS-2 in Figure 5). Maintenance and authenticating the medical knowledge becomes the responsibility of the KMS staff and IT specialists in intelligent medical network (IMS). Patient and physician databases are monitored and remain behind fire-walls for data, network and patient security. Access to outside medical service providers (MSPs) may be provided by specialized a secure vendor services network or by a secure Internet transactions. Such secure transactions are now carried by banking and financial networks. Much like the designs of MANs and WANs intelligent medical networks (IMNs) can be tailored to suit the need of any medical profession, any specialty and any community to serve patients and doctors, anywhere, anyplace and anytime.

Every sub-procedure is executed via an individual packet command (like the SS7, X.25 packet commands in the backbone network embedded in the intelligent networks). The net result of the procedure is conveyed to the users (or their application programs or APs) by a series of packet transactions. Such transactions between a single intelligent medical system and multiple knowledge banks are systematically processed, and the output is accumulated from sub-procedures, procedures, and runs. The entire usage of the network-based intelligent medical systems is as orderly and systematic as job processing in distributed computer environments. Debugging of this type of intelligent medical systems function becomes easy as the study of packet contents of any given procedure.

---

[5] Some of these concepts dealing with service "control point" (SCP), service "management system" (SMS), and network partitioning are borrowed from typical designs of various intelligent networks (US, French, Japanese, Swiss, etc.) and now in use by (almost) all communication and mobile networks around the world.

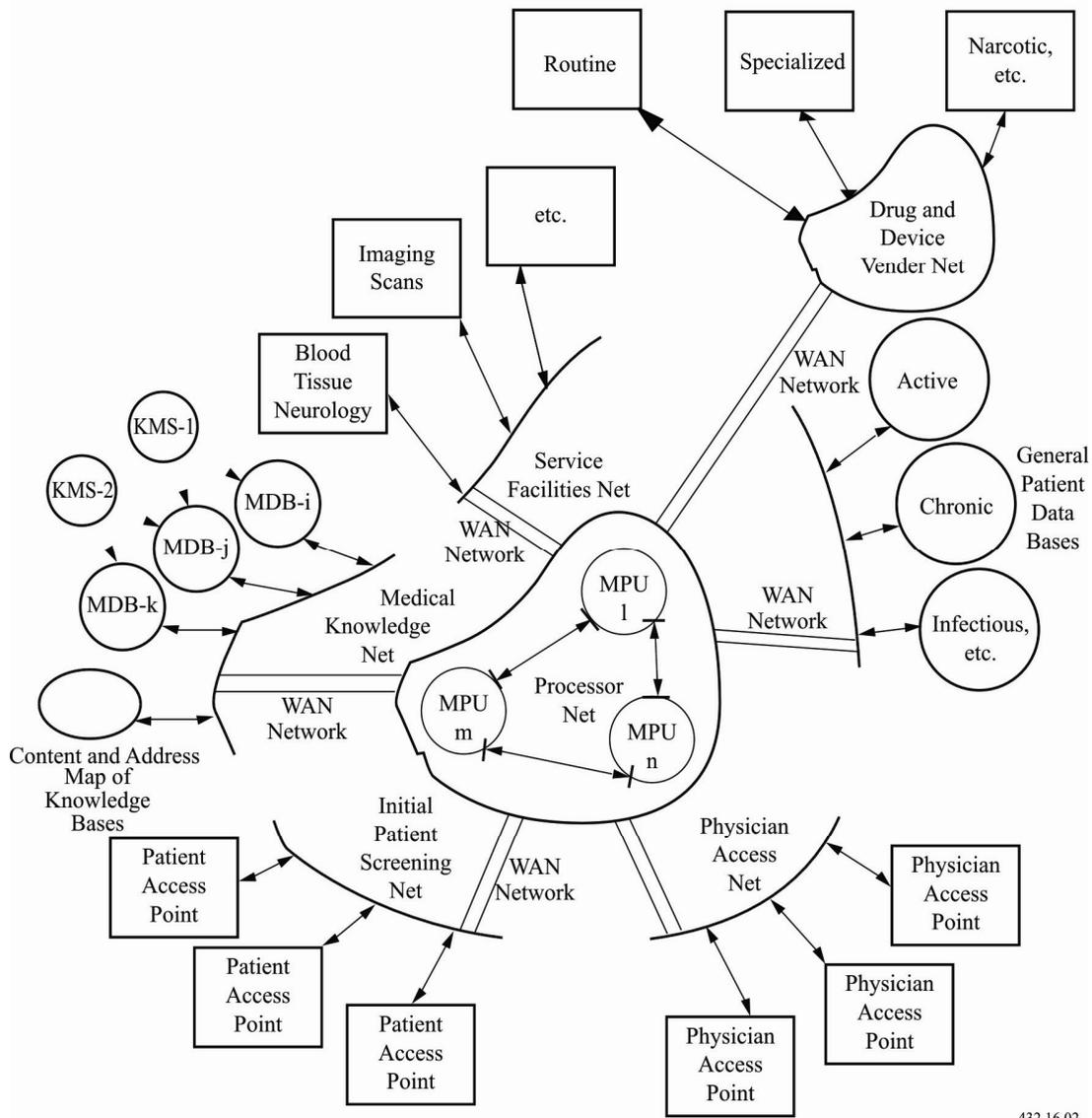

*Figure 5. Architecture of a WAN based intelligent medical network. A cluster of MPUs shares the overall task of running the medical facilities of any regional area. Existing hospitals and medical centers are networked to use their facilities optimally and economically. The resource allocation for the most expensive facilities is done by the operating system driving the processor net. Knowledge processing is also done by the processor net, thus sharing its own resources. KMS = knowledge machine(s); MDB = medical data base; MPU = medical processor unit; WAN = wide area network that facilitates resource sharing from numerous participating centers and hospitals.*

## 4. CONCLUSIONS

We evolve and present the architecture of a medical processor unit (MPU) in this paper. The platform for MPU design is based on the designs of classical graphics processor unit (GPU), the object processor unit (OPU) and the knowledge processor unit (KPU). Such specialized processors are being pursued by movie makers, games/graphics chip designers and designers of intelligent educational, instructional and knowledge based systems. Each of these specialized

26

application environment calls for optimality of architectural design and chip layout. In the OPU, KPU and MPU architectures presented here, numerous functions specific to the application are all built in the processor architecture. However, it is also possible to fragment the most frequently used and complex functions into modular arrays of simpler functions and then recombine them at a systems level.

We also provide a basis for building a medical machine (MM) and intelligent medical networks (IMNs) by combining the architectures of the MPU with the object memory modules, the switch, and the bus configuration(s) for object flow and operational code control signals. The IMN components and architectures can be highly customized to the functional and geographical distributions of medical facilities in a campus, region, nation or the globe. In this paper, we have delineated the issues and offered generic methodologies to address and resolve the issues specific to the medical environments.

The innovations and achievements in network design during the last two decades have made the dreams of communication designers a reality now. Looking back, we propose a migration to IMNs by (conceptually) replacing the electronic switching systems (AT&T's, ESS, that also process data and information) with medical machines and by (functionally) replacing the massively parallel processed 32-bit communication processor (3B2) with medical processors. The blending of medical, AI and communication processes is always the purgative of the next generation intelligent medical network designers.

## Authors' Short Biography

### Syed V. Ahamed

Syed Ahamed is a visiting professor in the Department of Computer Science and Engineering at the University of Hawaii, Hilo and a Professor of Medical Informatics, University of Medicine and Dentistry, Newark, NJ. Dr. Ahamed obtained his Ph.D. from Victoria University of Manchester in the United Kingdom. Dr. Ahamed also holds the Doctor of Science (D.Sc.) degree from Victoria University of Manchester as well as an M.B.A. in economics from New York University.

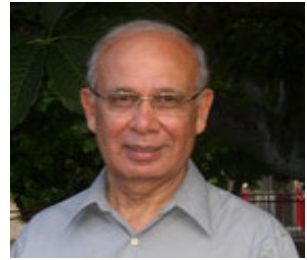

Dr. Ahamed, a Fellow of IEEE, has a broad spectrum of teaching and research interests including, Intelligent Sensing Systems, Computer Architecture, CPU and KPU Designs, Telecommunications, Intelligent Networks, Knowledge Systems, Management Information Systems, and Economics. He has received many European and US patents and published many books and hundreds of articles.

### Syed (Shawon) Rahman

Syed (Shawon) Rahman is an Assistant Professor in the Department of Computer Science & Engineering at the University of Hawaii-Hilo. Dr. Rahman's research interests include Software Engineering Education, Data Visualization & Scientific Computation, Information Assurance & Security, Web Accessibility, and Software Testing & Quality Assurance. He has published more than 50 pair-reviewed papers. He is an active member of many professional organizations including ACM, ASEE, ASQ, IEEE, and UPE.

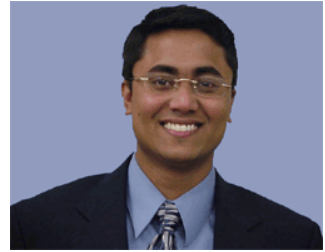